\documentstyle[epsfig,aps,prl,epsf,multicol,amssymb]{revtex}
\draft
\begin{document}
\hsize\textwidth\columnwidth\hsize
\csname@twocolumnfalse\endcsname

\title{Mode locking of vortex matter driven through mesoscopic channels}
\author{N. Kokubo$^1$, R. Besseling$^1$, V. M. Vinokur$^2$ and P. H. Kes$^1$\\}
\address{$^1$Kamerlingh Onnes Laboratory, Leiden University,P.O.Box 9504, 2300 RA Leiden, The
Netherlands.\\
$^2$Argonne National Laboratory, Materials Science
Division, 9700 South Cass Avenue, Argonne, IL 60439.}
\date{\today}
\maketitle

\begin{abstract}
We investigated the driven dynamics of vortices confined to
mesoscopic flow channels by means of a dc-rf interference
technique. The observed mode-locking steps in the $IV$-curves
provide detailed information on how both the number of vortex rows
and the lattice structure in each flow channel change with
magnetic field. Minima in flow stress occur when an integer number
of rows is moving coherently, while maxima appear when incoherent
motion of mixed $n$ and $n\pm 1$ row configurations is
predominant. Simulations show that the enhanced pinning at
mismatch originates from quasi-static fault zones with misoriented
edge dislocations induced by disorder in the channel edges.
\end{abstract}

\begin{multicols}{2}
\narrowtext \noindent During the last decades, vortex matter in
superconductors has been widely recognized as model system for
investigating driven, periodic media in various pinning
potentials. One of the intriguing phenomena it can display is a
non-monotonous behavior of the depinning current density $J_c$
with magnetic field. The best known example of such behavior is
the so called peak effect occurring in weak pinning
superconductors near $H_{c2}$ \cite{colpin}. Non-monotonous
behavior of $J_c$ also occurs in superconductors with (artificial)
periodic pinning arrays due to the possibility of
(in)commensurability between the pinning structures and the vortex
arrays \cite{DaldiniPRL74}. In both cases, the change in $J_c$ is
accompanied by pronounced structural transitions in the static as
well as the driven "lattice". For periodic pinning, it was
predicted theoretically \cite{Martinoli} and confirmed by
experiments \cite{Harada} that defects due to incommensurability
can cause a drastic reduction of $J_c$. For the case of the peak
effect in presence of random pinning the consensus is that the
opposite happens: here plastic deformations cause an {\it
increase} in $J_c$. However, the role of defects in the plastic
flow regime above the pinning threshold remains largely unknown.

A powerful tool to obtain time averaged, small scale information
on the moving structure is to probe the dc current-voltage ({\it
I-V}) characteristics in presence of superimposed rf-currents
\cite{Martinoli,FiorySchmidHaug,Look}. When a vortex lattice moves
{\it coherently} at average velocity $v$ through a pinning
potential, a microscopically periodic velocity modulation
\cite{Troyanovski} is induced at the washboard frequency $f_{int}=
v/a$ and, generally, at integer multiples $qf_{int}$. Here, $a$ is
the lattice period {\em in the direction of motion}. In presence
of an rf-force with frequency $f$ these modulations get
mode-locked when $f_{int}/f=p/q$ where $p$, like $q$, is an
integer, giving rise to interference plateaus in the {\it I-V}
curves. The appearance of this phenomenon was shown to depend
sensitively on the structure of the moving lattice
\cite{KoltonHarris}.

\begin{figure}
\epsfig{file=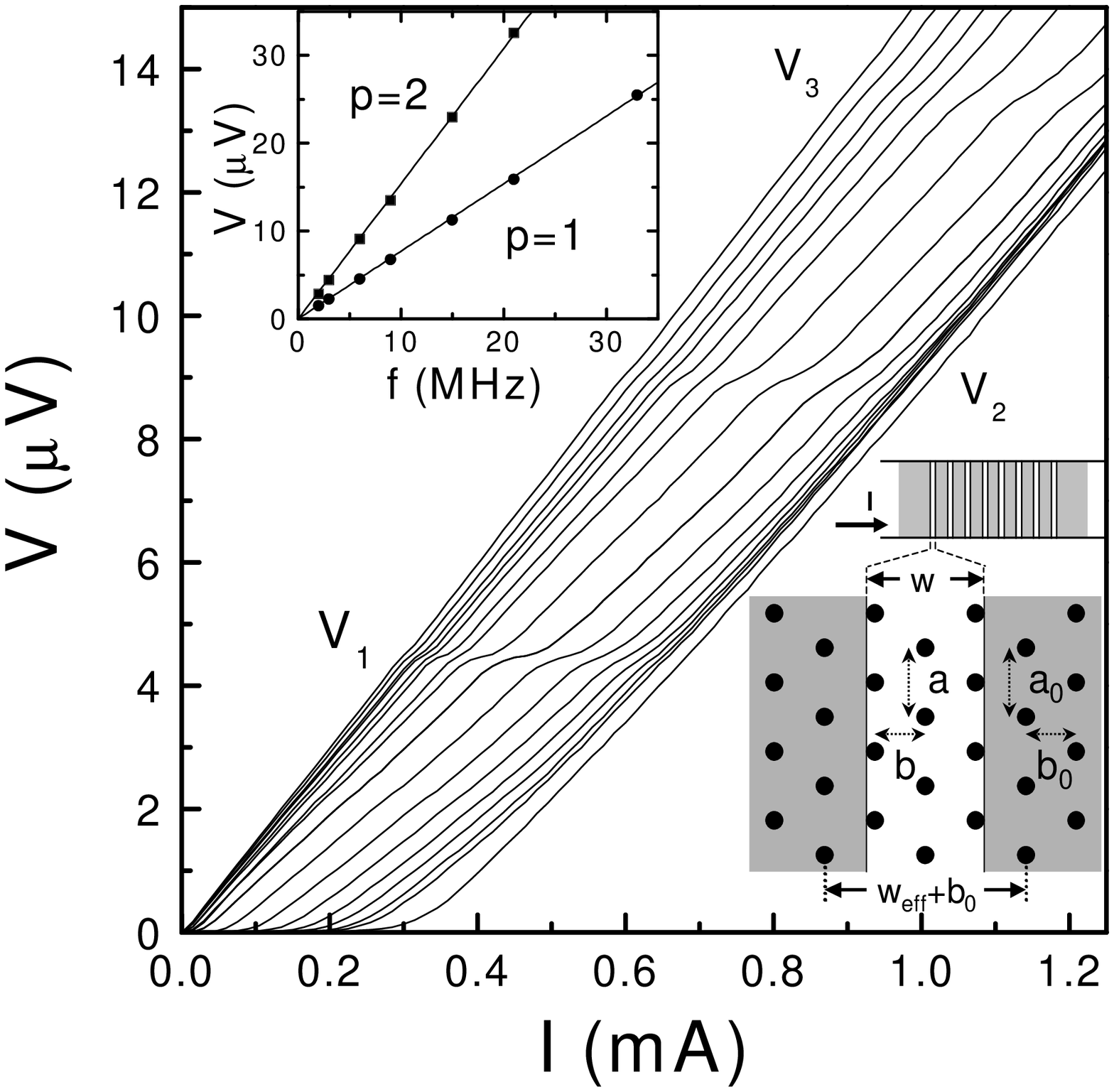, width=7cm, height=6.5cm} \vspace{0.4cm}
\caption{Main panel: dc-$IV$ curves at $60$ mT measured with
superimposed $6$ MHz rf currents (amplitude 4.7, 3.8, 3.0, 2.4,
1.9, 1.5, 1.2, 0.94, 0.75, 0.59, 0.47, 0.38, 0.30, 0.24, 0.19,
0.15 and 0mA from left to right). Interference voltages $V_p$ are
indicated. Upper inset: frequency dependence of the lowest two
interference voltages. Lower inset: schematic geometry of the
sample and a single channel. Strong pinning CE's are marked gray.
The (local) vortex lattice parameters and (effective) channel
width are indicated.} \label{figure1}
\end{figure}

In this Letter we report the results of mode-locking experiments
on vortex matter driven through mesoscopic channels. The channel
device \cite{Pruymboom} consists of an amorphous
(a-)Nb$_x$Ge$_{1-x}$ film ($x \simeq0.3$, $T_c=2.68$ K and
thickness $d=550$ nm) with a NbN film ($T_c=12$ K, $d=50$ nm) on
top. Straight parallel channels (width $w=230$ nm and length $300$
$\mu$m) were etched in the NbN layer. Pinning in a-NbGe is very
weak, while in NbN it is very strong \cite{fn_relat_Jc}, providing
easy vortex flow channels with strong pinning channel edges (CE's)
(see lower inset to Fig. \ref{figure1}). Importantly, the
structure of vortex matter in the channel can be tuned on the
'atomic' scale by changing, through the applied field, the
(mis)matching conditions between the lattice period and channel
width.

In our system the dc-critical current at which vortices in the
channel start to move is determined by the {\it (shear)
interaction with pinned vortices in the CE's}. Phenomenologically,
$J_c$ follows from $F_c=J_cB=2Ac_{66}/w_{eff}$ \cite{Pruymboom},
with $F_c$ the critical force density, $B$ the induction, $c_{66}$
the vortex lattice shear modulus and $w_{eff}$ the effective
channel width. The parameter $A$ depends crucially on the
microscopic structure of the vortex arrays. In fields for which
the arrays in the channel and the CE's are commensurate ($a=a_0$
and $w_{eff}/b_0$ integer in the inset to Fig. \ref{figure1} with
$a_0=2b_0/\sqrt{3}=(2\Phi_0/\sqrt{3}B)^{1/2}$ the equilibrium
lattice spacing), one expects a large flow stress, whereas it
should be small for incommensurate fields due to the occurrence of
misfit dislocations \cite{RutPRL99}. This picture seems correct,
since $J_c$ shows oscillations and is globally proportional to
$c_{66}$ \cite{Pruymboom}. However, recent studies \cite{Rut}
indicate that this picture is very sensitive to positional
disorder of the vortex arrays in the CE's, particularly when this
disorder is sufficient to cause spontaneous formation of
dislocations. Recent STM experiments \cite{Baarle} on vortices in
NbN indeed show strong deviations from the regular lattice
arrangement.

This new ingredient turns out to have intruiging consequences. By
means of the mode-locking technique we were for the first time
able to determine unambiguously the number of moving vortex rows,
the evolution of the microstructure and, from simulations, the
microscopic origin of the flow-impedance oscillations in the
channels.

The measurements were performed in a conventional four-probe
geometry. The rf-current was applied through an rf-transformer
with balanced transmission lines and a matching circuit. We used
frequencies from $1$ to $200$ MHz and amplitudes $I_{rf}<5$ mA. To
avoid heating, the sample was immersed in superfluid $^4$He. All
data shown here was obtained at $T=1.9$ K. Since strong
field-history effects exist \cite{Pruymboom}, we focus for
simplicity on results of field down measurements where first a
field $\sim \mu_0H_{c2}$ of NbGe is applied after which we slowly
($\mu_0dH/dt\simeq -2$ mT/s) sweep to the field under
consideration.

Figure \ref{figure1} shows a series of $IV$ curves measured at
$60$ mT with superimposed $6$ MHz rf-currents of various
amplitudes. The dc-$IV$ curve ($I_{rf}=0$) shows a nonlinear
voltage upturn from which $I_c$ was determined using a $100$ nV
criterion. The application of sufficiently large rf-currents leads
to clear mode-locking steps at equidistant voltage levels $V_p$
which are indicated for $p=1,2,3$. As shown below, $q=1$ and thus
$p=1$ refers to the fundamental and $p=2,3$ to higher harmonics.
These voltage levels depend linearly on $f$, as follows for
$p=1,2$ from the upper inset to Fig. \ref{figure1}. The
interference step width $\Delta I$ for each harmonic shows Bessel
function-like oscillations with $I_{rf}$ which will be discussed
in detail elsewhere.

\begin{figure}
\epsfig{file=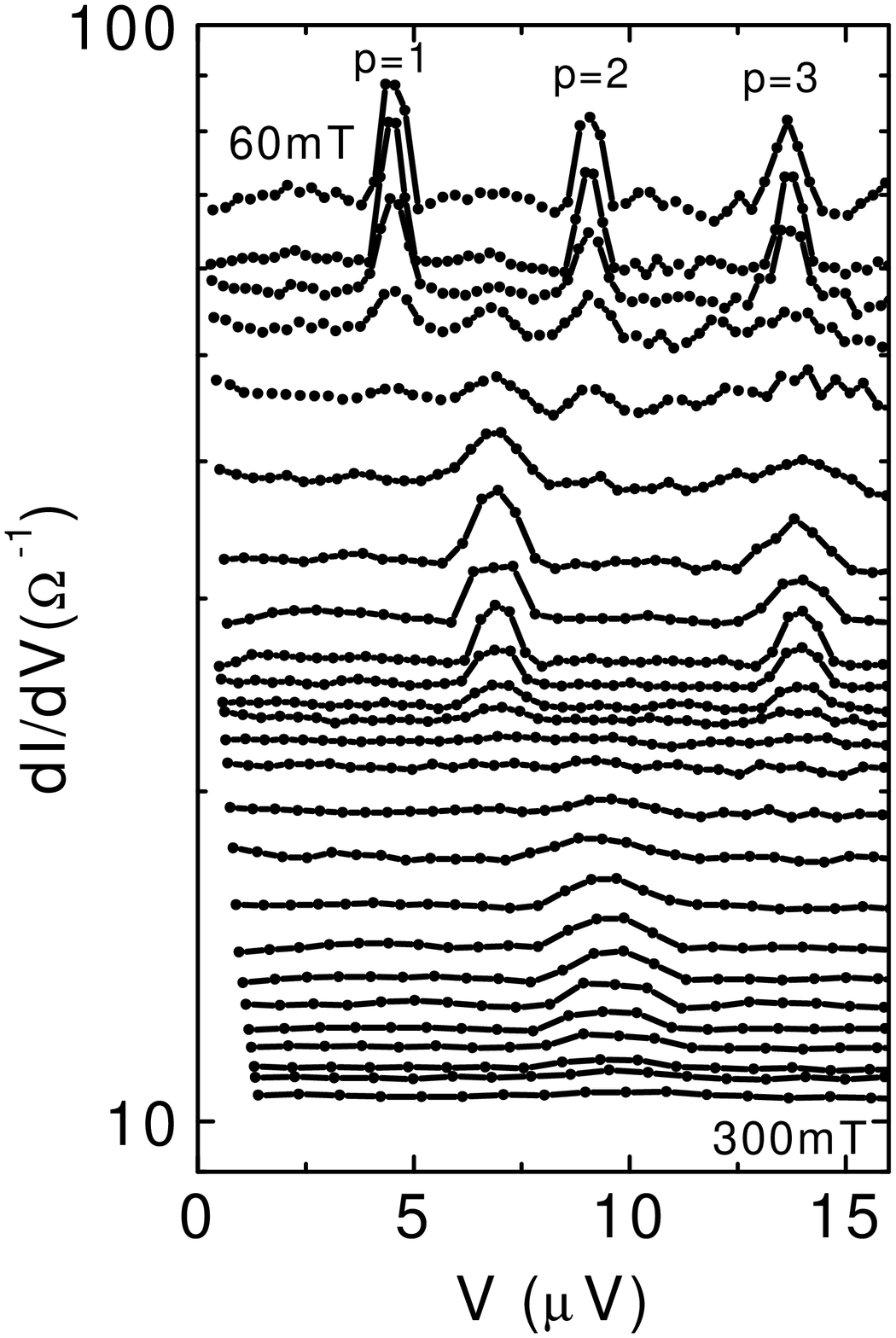, width=6.5cm, height=8.5cm}
\vspace{0.4cm} \caption{Differential conductance $dI/dV$ versus
$V$ as obtained from the $IV$ curves (as in Fig. \ref{figure1}):
the applied field is varied from $300$ mT (bottom) to $60$ mT
(top) in steps of $10$ mT. The amplitude of the superimposed $6$
MHz rf-current is $I_{rf}/I_c \sim 10$.} \label{figure2}
\end{figure}

Next we turn to the field dependence of the mode-locking
phenomenon. The characteristics are better displayed by plotting
the differential conductance $dI/dV$ versus $V$, as done in Fig.
\ref{figure2} for the $f=6$ MHz data. The results shown are
obtained at constant $I_{rf}\simeq 10 I_c$ in fields from $300$ mT
(bottom) to $60$ mT(top) with steps of $10$ mT. The interference
steps, now appearing as peaks in $dI/dV$, change in amplitude and
in voltage position with field. Focusing on the fundamental peak
at $V_1$, it is seen that in certain field intervals $V_1$ is
field independent, e.g. from $60-100$ mT $V_1$ is $4.6$ $\mu$V. At
$90$ mT another fundamental peak appears at a voltage of $6.9$
$\mu$V, which lies between the $p=1$ and $p=2$ interference peaks
in lower fields. The second harmonic now coincides with the $p=3$
peak at 60 mT. Between 100 and 110 mT, the lower voltage $p=1$
peak vanishes while the $6.9$ $\mu$V peak (and its higher
harmonics) grow and remain detectable up to $\approx 180$ mT.

\begin{figure}
\epsfig{file=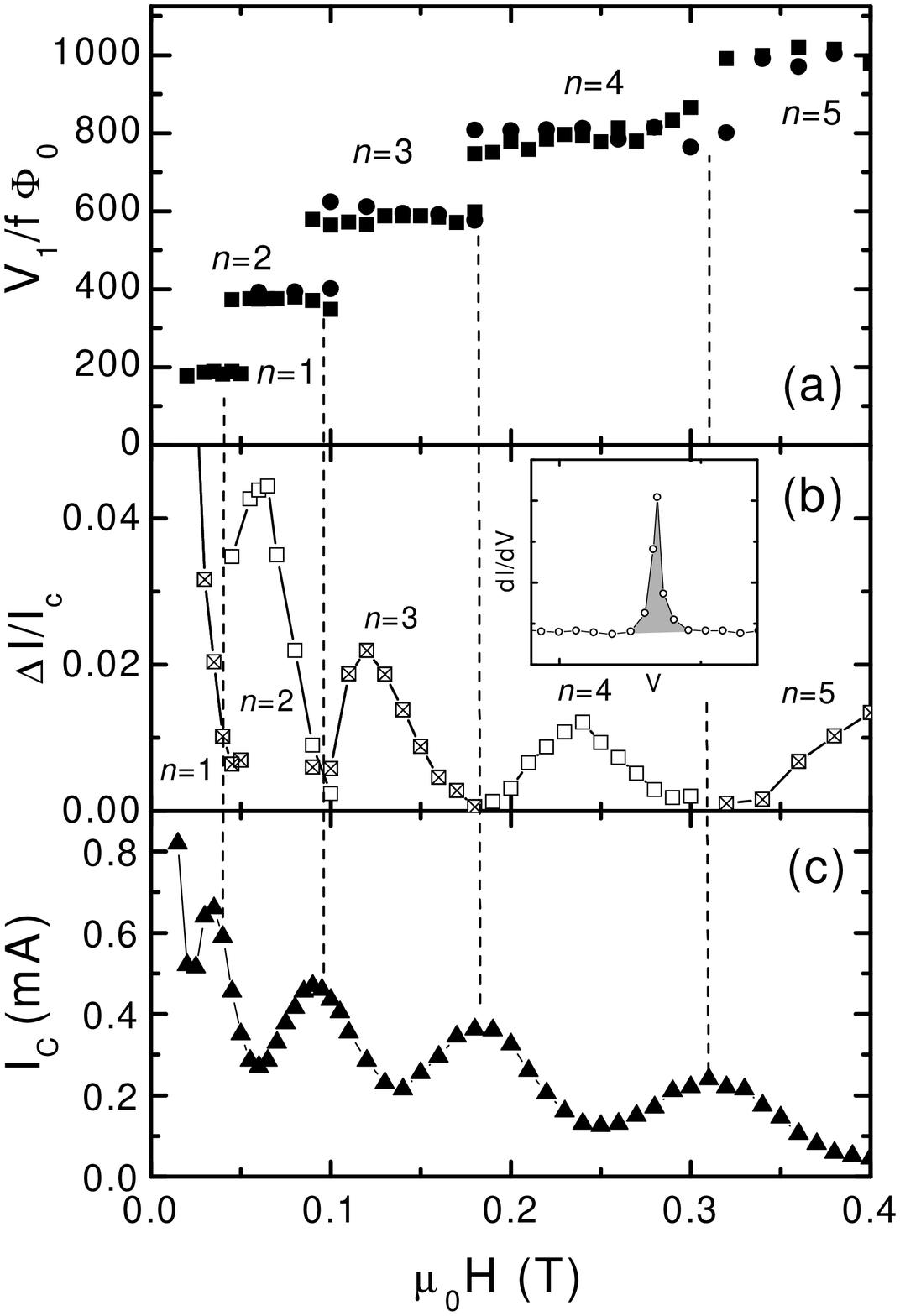, width=7cm, height=9.5cm} \vspace{0.4cm}
\caption{(a) $V_1/f$ in units $\Phi_0$ versus magnetic field
$\mu_0H$ for $6$ MHz ($\bullet$) and 60 MHz ($\blacksquare$). (b)
The normalized width of the interference current steps $\Delta
I/I_c$ of the $V_1$ peaks at $6$ MHz. Data belonging to odd $n$
are cross-marked for clarity. $\Delta I$ is defined by the gray
area in the inset. (c) The field dependence of the depinning
current $I_c$ determined from a $100$ nV criterion.}
\label{figure3}
\end{figure}

The described sequence of appearances and disappearances of the
peaks continues for larger fields. The resulting variation of
$V_1$ with field is shown in more detail in Fig. \ref{figure3}(a).
We plot the ratio of $V_1$ over $f$ to also include data taken at
$f=60$ MHz. Both data sets collapse on distinct voltage plateaus
separated by abrupt steps. Around the steps two consecutive
fundamental interferences can coexist. This behavior is
independent of frequency and $I_{rf}$. The voltage separation
between adjacent plateaus corresponds to the value of $V_1/f$ of
the lowest plateau. Therefore we labelled the plateaus by integers
$n$. This staircase-like behavior suggests that $V_1/f$ reflects a
process in which the number of vortex rows in each channel changes
with field in a discrete manner.

The relation between interference voltage and number of moving
rows can be derived as follows: the general interference condition
for the velocity of an {\em ordered} array with {\em average
longitudinal vortex spacing} $a$ and row spacing $b$ (see inset to
Fig. \ref{figure1}) is $v_{p,q}=(p/q)f a$. Thus the interference
voltage per channel $\tilde{V}_{p,q}=w_{eff}Bv_{p,q}$, with
$B=\Phi_0/ab$ the induction, is:
\begin{eqnarray}
\tilde{V}_{p,q} = \frac{p}{q}\Phi_0 f w_{eff}/b=\frac{p}{q}\Phi_0
f n,
\label{Vpqrelation}
\end{eqnarray}
where $w_{eff}/b=n$ is {\em the number of moving rows}. Taking
into account that we measure the voltage $V$ over $\sim 200$
channels, the height of the jumps in $V_1/f$ in Fig.
\ref{figure3}(a) corresponds to $\Phi_0$, validating the
identification of the peaks at $V_1$ as fundamental. We note that,
regardless of field, no subharmonic peaks ($q \neq 1$) could be
resolved. Such subharmonics {\it have} been observed in dc-rf
driven charge density waves and bulk vortex lattices
\cite{FiorySchmidHaug,Gruner}. Here, their absence might be
explained by the fact that our pinning potential, which stems from
the interaction with pinned vortices in the CE's with average
spacing $\sim a_0$, predominantly contains modes with wave vector
$\lesssim 2\pi/a_0$ \cite{Rut}.

The data in Fig. \ref{figure3}(a) combined with Eq.
(\ref{Vpqrelation}) shows unambiguously that in those parts of the
flow where vortices show mode locking, the number of moving rows
$n$ changes with field by one. We consider now in more detail the
field dependence of the amplitude $\Delta I$ of the mode locked
steps, i.e. the area under the interference peaks (see the inset
to Fig. \ref{figure3}(b)). Figure \ref{figure3}(b) shows $\Delta
I$ associated with the voltage plateaus for consecutive $n$ in
(a), normalized by $I_c$. At fields around the middle of the
plateaus, $\Delta I/I_c$ is large. This indicates a large degree
of coherency in the motion and implies that a unique, integer
number of rows $n$ is moving, corresponding to a matching
configuration as in the lower inset to Fig. \ref{figure1}.
Approaching the edges of the plateaus, $\Delta I/I_c$ decreases,
signalling a reduction of the coherency in the moving arrays. At
the plateau edges, $\Delta I/I_c$ is strongly suppressed, even if
$I_{rf}$ is varied. The coexistence of a residual signal from $n$
and $n\pm 1$ row configurations at the transition fields evidences
that these fields correspond to the maximum mismatch situation.

It is clear that upon approaching a mismatch condition, a moving
$n$ row configuration experiences increasing lattice strains which
will progressively induce misfit dislocations in the structure.
Let us now compare the observed transitions to the oscillations in
$I_c$ in Fig. \ref{figure3}(c). It is seen that {\em maxima in
$I_c$ appear at the $n \rightarrow n\pm 1$ transition fields},
while the minima in $I_c$ are located near matching fields
\cite{fn_othersample}. Therefore we conclude that, upon increasing
the density of misfit dislocations in the channel, $I_c$ is
enhanced. This result contrasts the picture in which dislocations
lower the flow stress and is strongly reminiscent of the mechanism
expected for the traditional peak effect in a random potential.

As mentioned, the positional disorder of vortex arrays in the
CE's, which is quenched for fixed field, plays a crucial role for
understanding the relation between mismatch and the dc-depinning
current $I_c$. We address this issue now in more detail by
molecular dynamics simulations \cite{RutPRL99,Rut} of dc-driven
channel vortices in the geometry of the lower inset to Fig.
\ref{figure1}. Vortex interactions were modelled by the London
potential with $\lambda/a_0=2$ ($\lambda$ is the penetration
depth) \cite{fnlama0insens}. The CE vortices were assigned random
shifts ${\bf d}$ with respect to the ideal lattice configuration,
such that $\sqrt{\langle(\nabla\cdot{\bf d})^2\rangle}\simeq
0.12$. We keep the average orientation of the CE arrays with a
principal axis parallel to the channel. The vortex density in the
channel and the CE's were both $(a_0b_0)^{-1}$, while the matching
condition was tuned by the ratio $w_{eff}/b_0$.

\begin{figure}
\epsfig{file=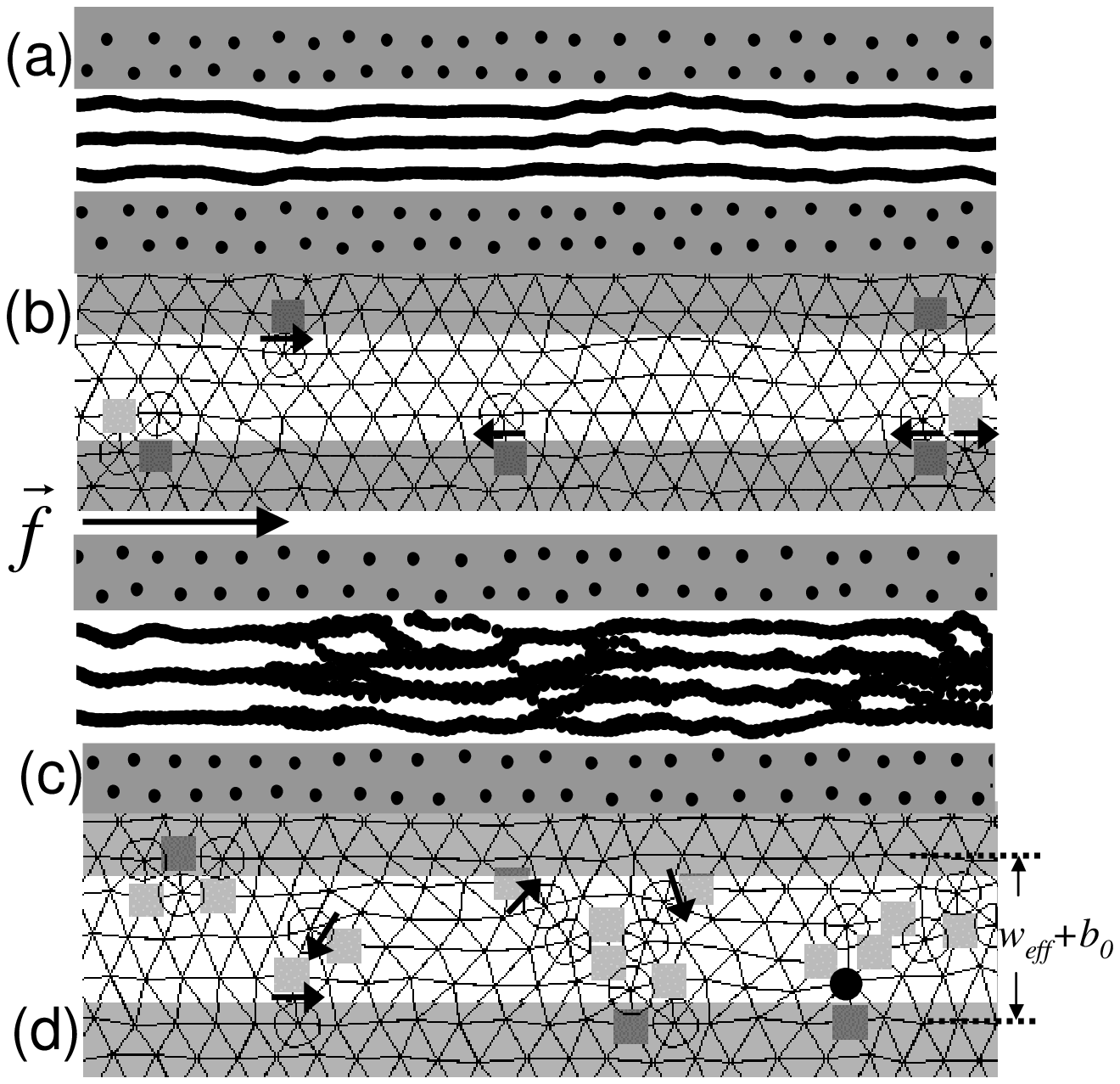, width=7cm, height=6.5cm} \vspace{0.4cm}
\caption{(a) Flow trajectories during motion over $5a_0$ around
matching ($w_{eff}/b_0=3.05$) for $J\simeq 3J_c^0$. Note the
positional disorder in the CE arrays. (b) Delauney triangulation
of one snapshot of (a). Direction of the drive
$\vec{f}=\vec{J}\times \vec{\Phi}_0$ is shown and Burgers vectors
of the dislocations are indicated by arrows. (c) Trajectories at
mismatch ($w_{eff}/b_0=3.55$) for $J\simeq 3.5 J_c^0$. (d)
Triangulation of a snapshot of (c). All data were taken for a
channel of length $L=150a_0$ and cyclic boundary conditions.}
\label{figure4}
\end{figure}

First we discuss the results of simulations around matching of a
$3$ row configuration ($w_{eff}/b_0=3.05$). The (rescaled) $J_c$
obtained in such simulations attains a value
$J_c=2Ac_{66}/(Bw_{eff})\equiv(A/A^0)J_c^0$ with $A \simeq 0.2A^0$
\cite{fnsimexp} where $A^0=(\pi\sqrt{3})^{-1}$ is the theoretical
value for shear flow of perfectly matching vortex arrays along
ordered CE's \cite{RutPRL99}. To understand the origin of this
reduced flow stress we turn to Fig. \ref{figure4}(a) and (b). In
(a) we show vortex trajectories during motion over $5a_0$ for a
dc-drive $J\simeq 3J_c^0$. The trajectories are essentially
parallel to the CE's and clearly show motion of $3$ rows. A
Delauney triangulation (Fig. \ref{figure4}(b)) of a snapshot of
(a), shows that vortices in the channel are mostly sixfold
coordinated and elastically coupled. However, at the CE's,
dislocations are present with Burgers vectors (denoted by arrows)
along the flow direction. These dislocations, which are mainly
induced by disorder in the CE's, are responsible for the reduction
of $J_c$ with respect to $J_c^0$.

To study a mismatch situation we consider the case
$w_{eff}/b_0=3.55$. The critical current we find is characterized
by $A=0.065\simeq 0.4A^0$, indeed enhanced compared to matching.
Figure \ref{figure4}(c) shows the corresponding trajectories for
$J\simeq 3.5 J_c^0$. Here, one observes coexistence of $3$ and $4$
moving rows and in addition substantial transverse displacements
and switching of vortices between rows. Importantly, the $3$ and
$4$ row regions remain quasi-static during motion and their
location is determined by static disorder in the CE's. The
triangulation in (d) of a snapshot of the flow reveals numerous
dislocations {\em inside the channel}. In contrast to the matching
case, their Burgers vectors assume all possible orientations and
are mostly misoriented to the flow direction. The motion of these
misoriented dislocations is blocked by the interaction with
disorder in the CE's and with neighboring misoriented
dislocations. This results in quasi-static fault zones where
during the flow the dislocation pattern changes irregularly and
the vortex trajectories are jammed. These phenomena form the
mechanism underlying the enhanced flow impedance at mismatch.

The overall picture that has emerged from these simulations is
that on approaching a matching state, the fault zones gradually
heal, yielding a gradual decrease of the critical current.
Simultaneously the size of coherent $n$-row regions grows,
providing an explanation for the experimentally observed increase
in the rf-dc interference signal. On changing the field, vortex
dynamics in the channels thus exhibits a series of smooth
structural transitions from quasi-1D coherent motion to quasi-2D
disordered flow in a disordered potential.

In summary, we investigated the flow of vortices in mesoscopic
channels bounded by pinned vortices. The rf-dc interference
measurements yield unambiguous information on the field evolution
of the number of moving rows and coherency in the channel. The
depinning current oscillations exhibit maxima at mismatch
conditions. Simulations show that this behavior originates from
blocking of misoriented dislocations in quasi-static fault zones,
which gradually heal on approaching a matching state.

This work was supported by the 'Stiching voor Fundamenteel
Onderzoek der Materie' (FOM). N.K. was supported by JSPS. V.M.V.
was supported by the DOE Office of Science.

\end{multicols}

\begin{references}
\bibitem{fnlama0insens}This interaction yields the correct $c_{66}$.
The results did not change for a larger ratio $\lambda/a_0$.
\bibitem{fnsimexp}The values $0.2A^0$ at matching and $0.4A^0$ at
mismatch are in reasonable agreement with the experimental
critical current obtained from linear extrapolation of the
$IV$-curves to $V=0$ \cite{Rut}.
\end{references}
\end{document}